\documentclass[journal]{IEEEtran}
\usepackage{ifpdf}
\usepackage{bm}
\usepackage{cite}
\usepackage{amsmath,amsfonts, amssymb, amsthm}
\usepackage{algorithmic}
\usepackage{algorithm}
\usepackage{array}
\usepackage[caption=false,font=normalsize,labelfont=sf,textfont=sf]{subfig}
\usepackage{textcomp}
\usepackage{stfloats}
\usepackage{url}
\usepackage{verbatim}
\usepackage{graphicx}
\usepackage{cite}
\usepackage{ulem}
\usepackage[colorlinks=true, linkcolor=black, citecolor=black, urlcolor=blue]{hyperref}
\usepackage{booktabs}
\usepackage{nicefrac}
\usepackage{microtype}
\usepackage{subdepth}
\usepackage{color}
\usepackage{colortbl}
\usepackage{bigstrut}
\usepackage{pifont}
\usepackage{caption}
\usepackage{multirow}
\usepackage{multicol}
\usepackage{float}
\usepackage{wrapfig}
\usepackage{utfsym}
\usepackage{makecell}

\usepackage{inconsolata}
\usepackage{arydshln}
\usepackage{threeparttable}
\definecolor{lightgrayrow}{gray}{0.9}

\definecolor{c6}{HTML}{000000}
\definecolor{c7}{HTML}{FF0000}
\definecolor{c8}{HTML}{0000FF}
\definecolor{c5}{HTML}{000000}

\begin{document}

\title{SLAM-LLM: A Modular, Open-Source Multimodal Large Language Model Framework and Best Practice for Speech, Language, Audio and Music Processing}

\author{Ziyang Ma, Guanrou Yang, Wenxi Chen, Zhifu Gao, Yexing Du, Xiquan Li, Zhisheng Zheng, Haina Zhu, Jianheng Zhuo, Zheshu Song, Ruiyang Xu, Tiranrui Wang, Yifan Yang, Yanqiao Zhu, Zhikang Niu, Liumeng Xue, Yinghao Ma, Ruibin Yuan, Shiliang Zhang, Kai Yu, Eng Siong Chng, Xie Chen\thanks{
Ziyang Ma, Guanrou Yang, Wenxi Chen, Xiquan Li, Haina Zhu, Jianheng Zhuo, Zheshu Song, Ruiyang Xu, Yifan Yang, Yanqiao Zhu, Zhikang Niu, Kai Yu, and Xie Chen are with the X-LANCE Lab, School of Computer Science, MoE Key Lab of Artificial Intelligence Shanghai Jiao Tong University, Shanghai, China.
Zhifu Gao and Shiliang Zhang are with the Tongyi Lab, Alibaba Group, Hangzhou, China.
Yexing Du is with the Peng Cheng Laboratory, Guangdong, China.
Zhisheng Zheng is with the University of Texas at Austin, Texas, U.S.
Tianrui Wang is with the Tianjin University, Tianjin, China.
Liumeng Xue and Ruibin Yuan are with the Hong Kong University of Science and Technology, Hongkong, China.
Yinghao Ma is with the Queen Mary University of London, London, U.K.
Ziyang Ma and Eng Siong Chng are with the Nanyang Technological University, Singapore.
Xie Chen is also with the Shanghai Innovation Institute, Shanghai, China.

\textit{This work was supported by the National Natural Science Foundation of China  (No. U23B2018), Shanghai Municipal Science and Technology Major Project under Grant 2021SHZDZX0102 and Yangtze River Delta Science and Technology Innovation Community Joint Research Project (2024CSJGG01100).}

\textit{Corresponding author: Prof. Xie Chen. e-mail: chenxie95@sjtu.edu.cn}

\textit{Open source at \url{https://github.com/X-LANCE/SLAM-LLM}}
}}

\markboth{Journal of \LaTeX\ Class Files,~Vol.~14, No.~8, August~2021}%
{Shell \MakeLowercase{\textit{et al.}}: A Sample Article Using IEEEtran.cls for IEEE Journals}


\maketitle

\begin{abstract}
The recent surge in open-source Multimodal Large Language Models (MLLM) frameworks, such as LLaVA, provides a convenient kickoff for artificial intelligence developers and researchers. 
However, most of the MLLM frameworks take vision as the main input modality, and provide limited in-depth support for the modality of speech, audio, and music. 
This situation hinders the development of audio-language models, and forces researchers to spend a lot of effort on code writing and hyperparameter tuning. 
We present SLAM-LLM, an open-source deep learning framework designed to train customized MLLMs, focused on speech, language, audio, and music processing. 
SLAM-LLM provides a modular configuration of different encoders, projectors, LLMs, and parameter-efficient fine-tuning plugins. 
SLAM-LLM also includes detailed training and inference recipes for mainstream tasks, along with high-performance checkpoints like LLM-based Automatic Speech Recognition (ASR), Automated Audio Captioning (AAC), and Music Captioning (MC). 
Some of these recipes have already reached or are nearing state-of-the-art performance, and some relevant techniques have also been accepted by academic papers. 
We hope SLAM-LLM will accelerate iteration, development, data engineering, and model training for researchers. 
We are committed to continually pushing forward audio-based MLLMs through this open-source framework, and call on the community to contribute to the LLM-based speech, audio and music processing. 
\end{abstract}


\section{Introduction}

\label{sec:intro}
The rapid advancement of Large Language Models (LLMs) has catalyzed the development of multimodal learning systems that integrate various forms of input such as text, vision, as well as audio. 
While recent open-source frameworks, such as LLaVA~\cite{liu2023visual} and OpenFlamingo~\cite{awadalla2023openflamingo} have demonstrated remarkable capabilities in vision-language modeling, they remain largely vision-centric. 
These frameworks offer limited support for non-visual modalities, particularly in the areas of speech, audio, and music processing. 
This lack of native support presents significant challenges for researchers working on auditory tasks, often forcing them to retrofit vision-based systems through cumbersome adaptations, resulting in inefficiencies and fragmented development workflows. 

To address these limitations, we introduce \textbf{SLAM-LLM: a modular, open-source framework for Multimodal Large Language Models (MLLMs), with a specific focus on speech, audio, and music.} 
SLAM-LLM is designed to lower the barrier for constructing, training, and deploying LLM-based systems that process audio-related modalities. Its core architecture follows a clean encoder–projector–LLM modular design, enabling seamless customization of model components through YAML configuration files. The framework supports a wide range of pretrained encoders (e.g., Whisper~\cite{radford2023robust}, HuBERT~\cite{hsu2021hubert}, BEATs~\cite{chen2022beats}, MERT~\cite{li2023mert}), projection modules (MLP layers, CNN layers, Q-Former~\cite{BLIP2}), and LLM backbones (e.g., LLaMA~\cite{touvron2023llama1, touvron2023llama2}, Vicuna~\cite{chiang2023vicuna}, Qwen~\cite{bai2023qwen}). It also incorporates parameter-efficient fine-tuning (PEFT) strategies such as Low-Rank Adaptation (LoRA) and prefix-tuning, making it suitable for low-resource and domain-specific adaptation. 
SLAM-LLM's design also addresses common engineering challenges in multimodal modeling, such as handling variable-length inputs and efficient fine-tuning, while supporting rapid prototyping and reproducibility. 

Beyond architectural flexibility, SLAM-LLM provides a comprehensive suite of training recipes and pretrained checkpoints for a variety of key tasks: 
Automatic Speech Recognition (ASR), Speech-to-Text Translation (S2TT), Speech Emotion Captioning (SEC), Automated Audio Captioning (AAC), and Music Captioning (MC). 
Extensive experiments across these tasks reveal a series of key insights. 

In brief, SLAM-LLM makes several significant contributions: 
\begin{enumerate}
    \item A unified, flexible, and extensible framework for developing LLM-based models for audio-related modalities; 
    \item Extensive modular support for diverse encoders, projectors, and large language models; 
    \item A library of curated training and inference recipes across speech, audio, and music tasks; 
    \item State-of-the-art results across benchmarks, particularly in speech recognition and audio captioning; 
    \item Open-source implementation encouraging community collaboration and rapid innovation. 
\end{enumerate}

SLAM-LLM bridges a major gap in the current MLLM ecosystem and promotes the development for future progress in general-purpose audio-language models (ALMs). 
By open-sourcing the framework and accompanying recipes, SLAM-LLM invites the broader community to contribute to and accelerate innovation in this important yet limitedly developed area of multimodal AI. 

\section{Multimodal Large Language Model Framework}

The development of LLMs has catalyzed a rapidly expanding ecosystem of open-source toolkits designed to support the training, fine-tuning, and deployment of LLMs across a variety of modalities. 
In particular, the recent surge of interest in multimodal learning has motivated the emergence of several frameworks that extend LLMs beyond pure text understanding. 

Among these toolkits, a number of general-purpose LLM infrastructure frameworks have been proposed to facilitate model development and customization. Notably, LLaMA-recipes~\cite{llama-recipes}, an official collection of implementations and usage patterns for the LLaMA model family released by Meta, provides end-to-end support for both text-based and visual-language models. 
LLaMA-Adapter~\cite{zhang2023llama} introduces a parameter-efficient tuning strategy for adapting frozen LLaMA models to instruction-following and, in its extended version, visual modalities, by introducing a vision-language alignment mechanism. 
LLaMA2-Accessory~\cite{gao2023llama} further generalizes this line of work, offering a comprehensive toolbox for pretraining, fine-tuning, and inference across both unimodal and multimodal settings, with native support for integrating visual encoders such as CLIP~\cite{radford2021learning} and DINOv2~\cite{oquab2023dinov2}. 
In parallel, LLaMA-Factory~\cite{zheng2024llamafactory} provides a unified and extensible training interface for over 100 foundation models, including multimodal branches that support image, audio, and video tasks via task-specific adapters and fine-tuning templates. 
Additionally, LMFlow~\cite{diao2023lmflow} and LitGPT offer efficient solutions for finetuning and inference, focusing on memory optimization, scalability, and support for multimodal inputs. 

In contrast to these infrastructure-focused toolkits, another line of work has produced end-to-end multimodal LLMs designed specifically for vision-language models (VLMs). 
MiniGPT-4~\cite{zhu2023minigpt} demonstrates that aligning a pretrained visual encoder with a frozen language model using a lightweight projection layer enables powerful vision-language dialogue capabilities. LLaVA~\cite{liu2023visual} adopts a two-stage training strategy that combines visual feature alignment with instruction tuning, resulting in a highly capable open-source visual dialogue model. TinyLLaVA~\cite{zhou2024tinyllava} builds on this approach with a focus on model efficiency, achieving competitive multimodal performance with significantly reduced parameter counts. 
Meanwhile, OpenFlamingo~\cite{awadalla2023openflamingo} reimplements DeepMind’s Flamingo~\cite{alayrac2022flamingo} architecture for open access, supporting few-shot learning across sequences of interleaved image-text pairs and offering pre-trained models at various scales. 

However, most existing toolkits primarily target text or vision tasks, offering limited or no support for non-visual modalities such as speech, audio, and music. As a result, researchers in auditory domains—e.g., speech recognition, audio captioning, or music analysis—must often repurpose tools whose architectures and pipelines are tightly coupled to vision or text, requiring substantial adaptation effort. This lack of native support hinders progress in building efficient audio-language models. 
While toolkits like ESPnet\footnote{\url{https://github.com/espnet/espnet}} focus on end-to-end speech systems and Fairseq\footnote{\url{https://github.com/facebookresearch/fairseq}} focus on speech sequence modeling, they do not adopt an LLM-centric design, which is a paradigm shift in SLAM-LLM. The framework addresses this gap with a modular design that seamlessly integrates speech, audio, and music encoders with LLMs. In SLAM-LLM, all auditory tasks are unified into an auto-regressive generation process, allowing researchers to leverage the strong text power of LLMs while modularly incorporating proven speech processing best practices to guide the generative performance. 
Its flexible architecture supports easy customization of encoders, projectors, and LLMs, and is compatible with PEFT~\cite{peft}, FSDP~\cite{zhao2023pytorch}, and DeepSpeed~\cite{rasley2020deepspeed}. With comprehensive training and inference recipes, SLAM-LLM accelerates research and promotes community-driven development via its open-source release.

\section{Design of SLAM-LLM}

\begin{figure}[htbp]
\centering
\includegraphics[width=1\linewidth]{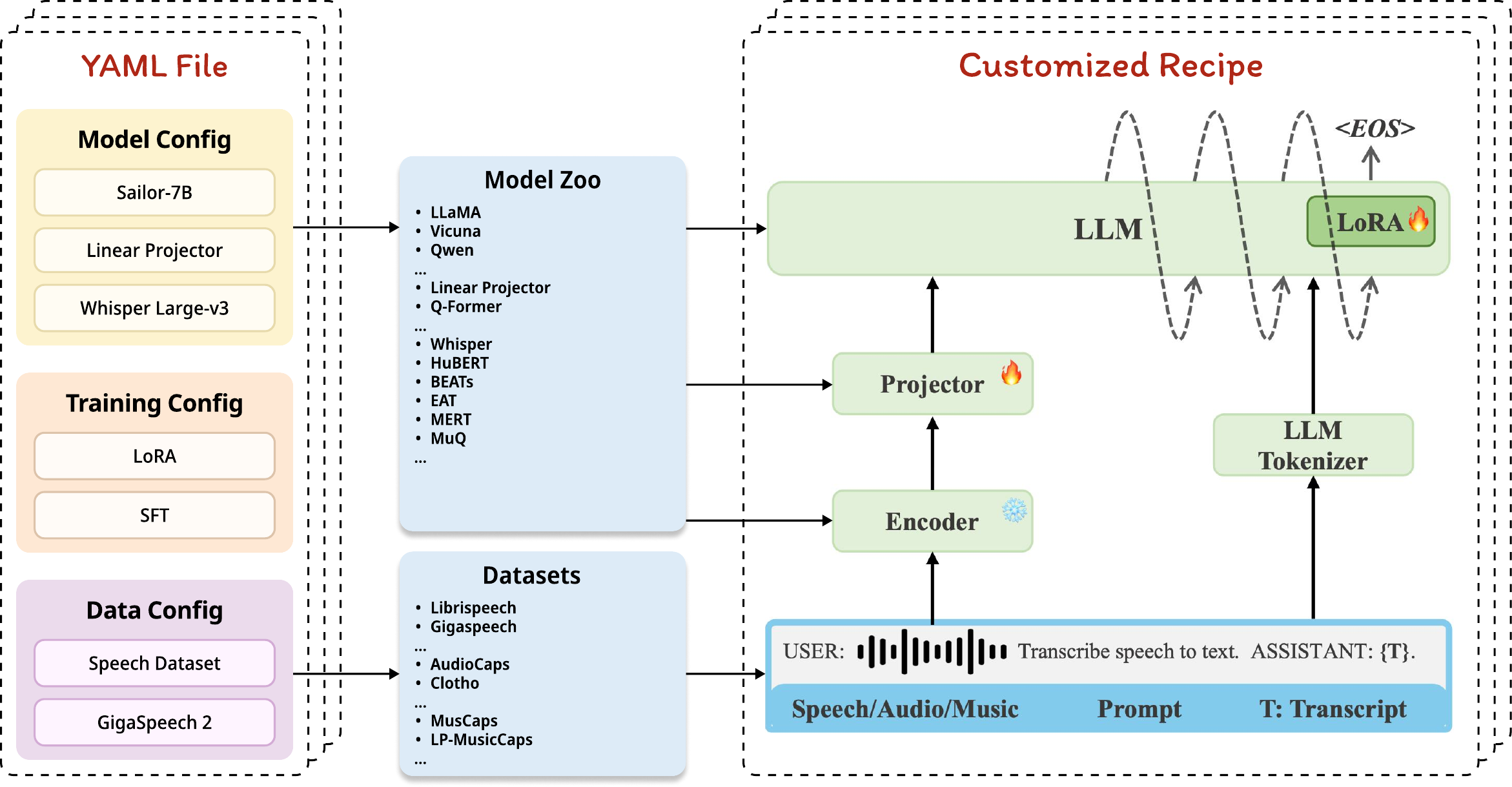}
\caption{A brief workflow in the SLAM-LLM framework. By specifying model, training, and data configurations in a YAML file, customized requirements are met efficiently. The depicted example demonstrates an LLM-based Southeast Asian ASR model using a Southeast Asian language LLM Sailor-7B and corresponding speech dataset GigaSpeech 2.}
\label{fig:pipeline}
\end{figure}

\subsection{Overview of SLAM-LLM}
SLAM-LLM allows users to configure customized MLLMs through a YAML file, significantly reducing the burden on model construction. 
Users specify the necessary model components, training strategies, and data formats in the YAML file, enabling SLAM-LLM to tailor the training and inference according to specific customized requirements. 

Figure~\ref{fig:pipeline} exemplifies how SLAM-LLM operates using an LLM-based low-resource language ASR model. 
Specifically, during training, the YAML file designates Whisper Large-v3~\cite{radford2023robust} as the speech encoder, a linear layer as the projector, and the Southeast Asian language model Sailor-7B~\cite{sailor} as the LLM, using the GigaSpeech 2~\cite{gigaspeech2} dataset for LoRA fine-tuning. This setup effectively configures an LLM-based Southeast Asian ASR model. 
For the trained model, only the parameters tuned (such as the projector and LoRA) need to be saved, and treated as generated assets. 
During inference, other components can be fetched from the Model Zoo and assembled along with generated assets according to the configuration, greatly enhancing flexibility in the process of development. 

\begin{figure}[htbp]
\centering
\includegraphics[width=1\linewidth]{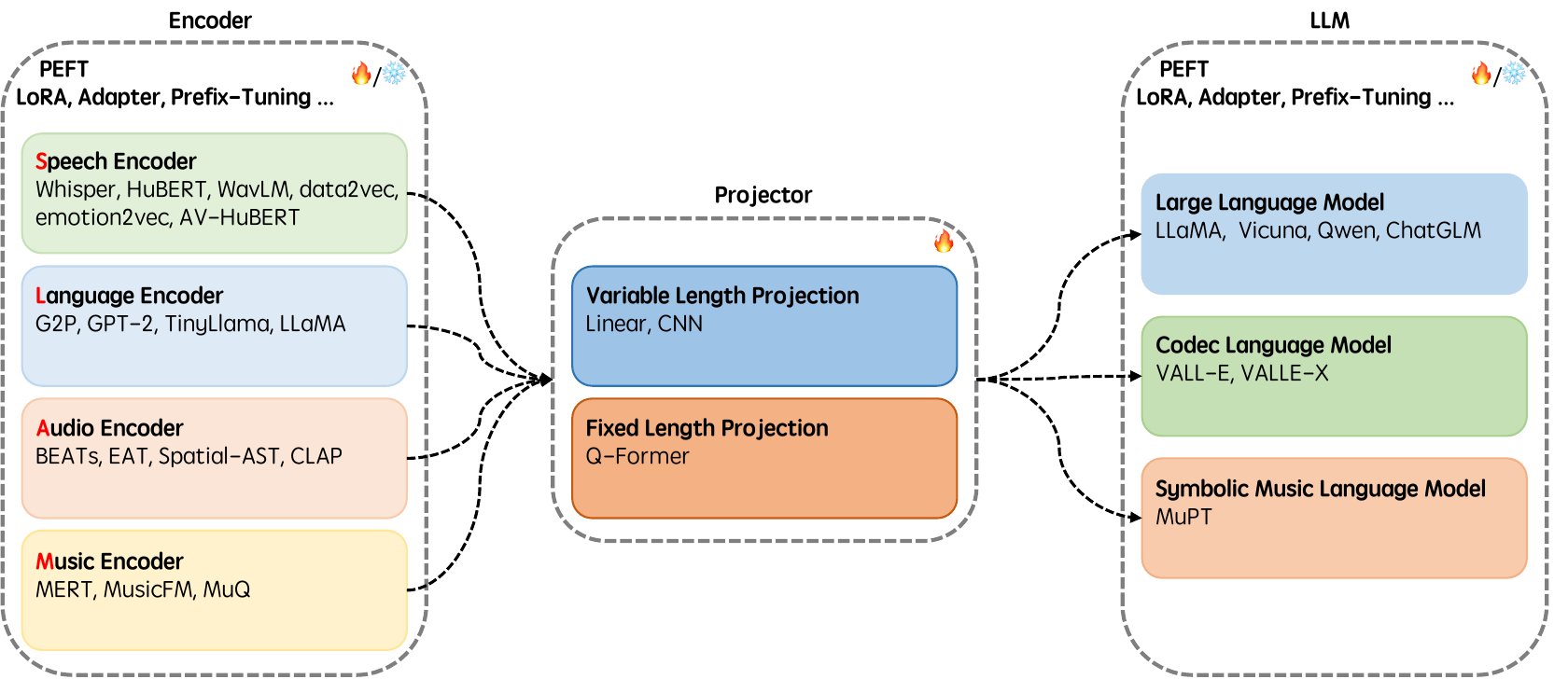}
\caption{SLAM-LLM modularizes the Encoder-Projector-LLM three-part components, which are assembled according to the configuration at training and inference time. }
\label{fig:supported}
\end{figure}

\subsection{Modular Design}
A distinguishing feature of SLAM-LLM is its modular design, which qualifies it as a framework rather than a mere model. 
As illustrated in Figure~\ref{fig:supported}, SLAM-LLM decomposes the MLLM into three decoupled components: Encoder, Projector, and LLM. 
Each component can be individually specified if and how to be tuned according to the training configuration. 

More specifically, the encoder serves as a perceiver of signals, supporting various types of encoders for speech, language, audio, and music. 
The projector bridges the encoder and the LLM. SLAM-LLM provides support for variable-length projection, such as the linear projector, as well as fixed-length projection like Q-Former. 
The LLM leverages its extensive knowledge and generation capabilities to produce high-quality outputs.
Within SLAM-LLM, diverse types of language models are supported, including text language models, codec language models, and symbolic music language models, catering to the customized needs in the fields of speech, audio, and music. 

Based on the modular design of SLAM-LLM, we have provided various recipes and checkpoints for implementing different tasks. 
In the following experiment sections, we present fundamental results for most classic tasks in detail, prioritize using mainstream model components wherever possible, and conduct large-scale experiments. 
By comparing and analyzing various experimental results, we derive several empirical insights. These insights not only guide the optimization of existing systems but also lay the foundation for developing new components. 
Furthermore, these results can serve as a robust baseline reference for future research and applications.

\subsection{Task Support}
While SLAM-LLM is primarily designed to facilitate LLM-based understanding tasks (i.e., mapping audio inputs to natural language text), it has been extended to support speech generation and multi-turn dialogue to further its utility as a general-purpose framework. For generation tasks, the framework supports codec-based text-to-speech (TTS) models such as VALL-E X~\cite{zhang2023speak}, leveraging its native support for codec language models. Furthermore, SLAM-LLM incorporated SLAM-Omni~\cite{chen2025slam}, an end-to-end, multi-turn speech dialogue system, processing continuous speech representations and generating discrete speech tokens as output. Such a design demonstrates the framework's flexibility in supporting complex, interaction-oriented applications that go beyond static tasks like automated audio captioning or speech recognition. 

\begin{table*}[htbp]
\centering
\caption{Different combinations of supervised speech encoders and LLMs to conduct LLM-based ASR.
We show results of Whisper models with different sizes on LLMs of different scales.}
\label{tab:asr_librispeech_supervised_encoder}
\begin{tabular}{l|cccc|cccccc}
\hline
\multirow{3}{*}{\textbf{Speech Encoder}} & \multicolumn{4}{c|}{\textbf{Pre-trained Model}} & \multicolumn{6}{c}{\textbf{Chat Model}} \\
& \multicolumn{2}{c}{TinyLLaMA-1.1B} & \multicolumn{2}{c|}{LLaMA-2-7B} & \multicolumn{2}{c}{TinyLLaMA-Chat-1.1B} & \multicolumn{2}{c}{LLaMA-2-Chat-7B} & \multicolumn{2}{c}{Vicuna-v1.5-7B} \\
& clean & other & clean & other & clean & other & clean & other & clean & other \\
\hline
Whisper-tiny &12.72 &21.64 &16.16 &25.17 &9.55 &21.01 &8.97 &18.77 & 7.07 & 16.01\\
Whisper-base &7.35 &15.89 &17.46 &21.84 &7.03 &15.92 &6.37 &12.98 & 5.07 & 13.07\\
Whisper-small &6.61 &11.81 &6.41 &10.88 &5.94 &11.5 &4.51 &8.94 & 4.19 & 9.50\\
Whisper-medium &4.65 &8.95 &3.35 &6.10 &5.01 &8.67 &2.71 &6.37 & 2.72 & 6.79\\
Whisper-large &4.39 &8.22 &3.01 &7.15 &4.33 &8.62 &2.72 &6.79 & 2.58 & 6.47\\
\hline
\end{tabular}
\end{table*}

\section{LLM-based Speech Processing}

\subsection{Automatic Speech Recognition (ASR)}

\paragraph{Model setup}
We investigate two types of speech encoders: supervised models trained on large-scale speech-text pairs and self-supervised models trained on unlabeled speech. For supervised encoders, we use the Whisper~\cite{radford2023robust} family (\textit{tiny} to \textit{large-v2}), retaining only the encoder as a feature extractor. For self-supervised encoders, we evaluate HuBERT~\cite{hsu2021hubert} and WavLM~\cite{chen2022wavlm} at multiple scales, including both pre-trained and fine-tuned versions. Base models are trained on the 960-hour LibriSpeech~\cite{panayotov2015librispeech}, while larger ones use LibriLight~\cite{kahn2020libri} (60k hours for HuBERT; 94k hours + VoxPopuli~\cite{wang2021voxpopuli} + GigaSpeech~\cite{chen2021gigaspeech} for WavLM). HuBERT also includes X-Large variants, among the largest publicly available.
For LLMs, we consider both pre-trained and chat-based variants with supervised fine-tuning and RLHF. The pre-trained models include \textit{TinyLLaMA} (1.1B)\cite{zhang2024tinyllama} and \textit{LLaMA-2} (7B)\cite{touvron2023llama2}; chat models include \textit{TinyLLaMA-Chat} (1.1B), \textit{LLaMA-2-Chat} (7B), and \textit{Vicuna} (7B)~\cite{chiang2023vicuna}.
All encoder outputs are at 50 Hz and downsampled to 10 Hz before feeding into the LLM. The projector uses a hidden size of 2048; encoder output and LLM input dimensions vary by model. The training recipes and model checkpoints can be found here\footnote{\url{https://github.com/X-LANCE/SLAM-LLM/tree/main/examples/asr_librispeech}}.
To address the practical requirements for real-world deployment, we report the compute resources for a representative task using a 7B-scale LLM backbone: while training configurations may vary slightly between experiments to ensure peak performance, the underlying resource requirements remain similar. Training an LLM-based model on 1,000 hours of speech data for 1 epoch, while keeping the LLM and encoder frozen and updating only the projector, requires approximately 4 hours using 4 × NVIDIA A100 (80GB) GPUs. 

\paragraph{Datasets}

We use the LibriSpeech~\cite{panayotov2015librispeech} benchmark with 960 hours of training data, without augmentation or splicing. Validation and testing are conducted on the 10-hour dev-other, test-clean, and test-other subsets. 

\paragraph{Experiments on different supervised speech encoders}

Table~\ref{tab:asr_librispeech_supervised_encoder} presents ASR results across various combinations of supervised speech encoders and LLMs, highlighting several key findings.
First, ASR performance improves with larger Whisper encoders. For example, using TinyLLaMA-1.1B, the test-clean WER drops from 12.72 (Whisper-tiny) to 4.39 (Whisper-large), showing that larger encoders better capture speech features. However, returns diminish with size: the WER improvement from Whisper-medium (4.65) to Whisper-large (4.39) is marginal.
Second, chat models consistently outperform pre-trained LLMs across all encoder sizes. For instance, with Whisper-large, LLaMA-2-7B yields a test-clean WER of 3.01, while LLaMA-2-Chat-7B improves to 2.72. This suggests SFT enhances the model’s ability to treat speech embeddings as a form of "language", benefiting from translation-like learning during fine-tuning.
Finally, with Whisper-large fixed, Vicuna-7B achieves the best results, reaching a 2.58 WER on test-clean—surpassing both LLaMA-2-7B (3.01) and LLaMA-2-Chat-7B (2.72). This indicates that Vicuna, fine-tuned on user-shared conversational data, generalizes more effectively.

\paragraph{Experiments on different self-supervised speech encoders}

Table~\ref{tab:asr_librispeech_self_supervised_encoder} analyzes self-supervised speech encoders for LLM-based ASR, using Vicuna-7B-v1.5 as the fixed LLM.
WER generally improves with larger encoder sizes, mirroring trends observed with supervised models. At the base scale (~95M parameters, 768 hidden size), HuBERT and WavLM perform on par with Whisper-small, showing no clear advantage for self-supervised learning. However, at 300M parameters, WavLM Large surpasses all supervised encoders in Table\ref{tab:asr_librispeech_supervised_encoder}, including Whisper-medium (300M) and Whisper-large (600M), while HuBERT’s gains from Base to Large are more modest. 
Fine-tuning HuBERT Large on LibriSpeech yields substantial improvements, achieving 2.10\% WER on test-clean and 4.26\% on test-other—outperforming WavLM Large. Scaling further to HuBERT X-Large (~1B parameters, fine-tuned) delivers the best results: 1.84\% WER on test-clean and 3.39\% on test-other, representing relative reductions of 28.7\% and 47.4\% over Whisper-large. These findings underscore the strong benefits of scaling and fine-tuning SSL encoders for LLM-based ASR.

\begin{table}[htbp]
\setlength\tabcolsep{1.5pt}
\centering
\caption{Explore the performance with different SSL speech encoders for LLM-based ASR. 
LS-960 means the Librispeech 960 hours dataset. FT donates Fine-tuning. 
}
\label{tab:asr_librispeech_self_supervised_encoder}
\resizebox{1\linewidth}{!}{
\begin{tabular}{lccccc}
\hline
\multirow{2}{*}{\makecell{ \textbf{Speech}\\\textbf{Encoder}}} &\multirow{2}{*}{\makecell{ \textbf{Encoder}\\\textbf{Params}}} &\multirow{2}{*}{\makecell{ \textbf{Hidden}\\\textbf{Size}}} & \multirow{2}{*}{\makecell{ \textbf{Projector}\\\textbf{Params}}}  & \multicolumn{2}{c}{\textbf{WER(\%) $\downarrow$}} \\
& & & & \textbf{test-clean} & \textbf{test-other} \\
\hline
HuBERT Base & 94.70M & 768 & 16.26M & 4.43 & 10.72 \\
WavLM Base & 94.38M & 768 & 16.26M & 4.14 & 9.66 \\
HuBERT Large & 316.61M & 1024 & 18.88M &  4.53 & 8.74 \\
\quad + LS-960 FT & 316.61M & 1024 & 18.88M & 2.10 & 4.26 \\
WavLM Large & 315.45M & 1024 & 18.88M & 2.13 & 4.73 \\
\quad + LS-960 FT & 315.45M & 1024 & 18.88M & 1.96 & 4.18 \\
HuBERT X-Large & 964.32M & 1280 & 21.50M & 2.41 & 4.49 \\
\quad + LS-960 FT & 964.32M & 1280 & 21.50M & 1.84 & 3.39 \\
\hline
\end{tabular}
}
\end{table}

\paragraph{Comparison with Non-LLM-based ASR models}

Table~\ref{tab:asr_librispeech_nn} compares our LLM-based ASR model (SLAM-LLM) with state-of-the-art NN-based systems, including specialist models trained on LibriSpeech-960, self-supervised models pre-trained on large-scale unlabeled speech, and universal models trained on multilingual labeled datasets.
For LibriSpeech-960 specialist models, we compare against ContextNet~\cite{han2020contextnet}, Conformer~\cite{gulati2020conformer}, Branchformer~\cite{peng2022branchformer} (ESPnet), and Zipformer~\cite{yao2023zipformer} (K2). These models use extensive system-level techniques—SpecAugment, speed perturbation, EMA averaging—and often fuse in-domain LMs trained on LibriSpeech text. Despite forgoing such engineering, our LLM-based model outperforms the best of these specialist systems. 
However, SSL models like HuBERT-large/x-large implemented with Fairseq and WavLM-large implemented with UniSpeech, pre-trained on larger unlabeled datasets and paired with in-domain LMs, outperform our model on the noisy test-other subset. These gains are largely attributable to LM integration, suggesting that domain-specific LLM fine-tuning may offer similar improvements, albeit at the cost of generalization.
Compared to general-purpose models, our system surpasses Whisper-large-v2, the stronger Whisper-large-v3, and OWSM-v3.1, a top academic baseline. These results highlight the effectiveness of our LLM-based approach, achieving competitive or superior ASR performance without heavy engineering or massive pre-training.

\begin{table}[htbp]
\centering
\setlength\tabcolsep{1pt}
\caption{
Compared to prior NN-based models on LibriSpeech, \textbf{\textit{Specialist Models}} are trained solely on the 960-hour LibriSpeech dataset, often using \textbf{\textit{in-domain LMs}} built from LibriSpeech text and transcripts. \textbf{\textit{SSL-based Models}} are self-supervised models pre-trained on large-scale unlabeled data, then fine-tuned with supervision. \textbf{\textit{Universal Models}} refer to general-purpose systems trained on massive labeled speech-text pairs. Repository links are provided for reproducibility. “Ours” denotes results from our best system in Table~\ref{tab:asr_librispeech_self_supervised_encoder}.
}
\label{tab:asr_librispeech_nn}
\resizebox{1\linewidth}{!}{
\begin{tabular}{lcccc}
\hline
\multirow{2}{*}{\textbf{Model}} & \multirow{2}{*}{\textbf{Implementation}} & \textbf{Speech Data(h)} & \multicolumn{2}{c}{\textbf{WER(\%) $\downarrow$}} \\
& & \textbf{Unlabeled/Labeled} & \textbf{test-clean} & \textbf{test-other} \\
\hline
\multicolumn{3}{l}{\textbf{\textit{Specialist Models}}} \\
\hline
ContextNet-large~\cite{han2020contextnet} & \multirow{2}{*}{-} & \multirow{2}{*}{-/960} & 2.1 & 4.6 \\
\quad + in-domain LM & & & 1.9 & 4.1 \\
Conformer-large~\cite{gulati2020conformer} & \multirow{2}{*}{-} & \multirow{2}{*}{-/960} & 2.1 & 4.3 \\
\quad + in-domain LM & & & 1.9 & 3.9 \\
Branchformer-large~\cite{peng2022branchformer} & \multirow{2}{*}{ESPnet} & \multirow{2}{*}{-/960} & 2.4 & 5.5 \\
\quad + in-domain LM & & & 2.1 & 4.5 \\
Zipformer-large~\cite{yao2023zipformer} & \multirow{2}{*}{K2} & \multirow{2}{*}{-/960} & 2.0 & 4.4 \\
\quad + in-domain LM & & & 1.9 & 3.9 \\
\hline
\multicolumn{3}{l}{\textbf{\textit{SSL-based Models}}} \\
\hline
wav2vec 2.0-large~\cite{baevski2020wav2vec} & \multirow{2}{*}{Fairseq} & \multirow{2}{*}{60k/960} & 2.2 & 4.5 \\
\quad + in-domain LM & & & 1.8 & 3.3\\
HuBERT-large~\cite{hsu2021hubert} & \multirow{2}{*}{Fairseq} & \multirow{2}{*}{60k/960} & 2.2 & 4.5 \\
\quad + in-domain LM & & & 1.9 & 3.3\\
HuBERT-x-large~\cite{hsu2021hubert} & \multirow{2}{*}{Fairseq} & \multirow{2}{*}{60k/960} & 2.0 & 3.7 \\
\quad + in-domain LM & & & 1.8 & 2.9\\
WavLM-large~\cite{chen2022wavlm} & \multirow{2}{*}{UniSpeech} & \multirow{2}{*}{94k/960} & 2.7 & 5.0 \\
\quad + in-domain LM & & & 1.8 & 3.2\\
\hline
\multicolumn{3}{l}{\textbf{\textit{Universal Models}}} \\
\hline
Whisper-large-v2~\cite{radford2023robust} & OpenAI/Whisper & -/680k & 2.7 & 5.2 \\
Whisper-large-v3~\cite{radford2023robust} & OpenAI/Whisper & -/1000k & 2.0 & 3.9 \\
OWSM-v3.1~\cite{peng2024owsm} & ESPnet & -/180k & 2.4 & 5.0 \\
\hline
\multicolumn{3}{l}{\textbf{\textit{LLM-based Models}}} \\
\hline
Ours & SLAM-LLM & -/960 & 1.8 & 3.4 \\
\hline
\end{tabular}
}
\end{table}

\paragraph{Experiments on minority language datasets}

Beyond high-resource languages like English and Chinese, low-resource speech recognition remains a major challenge. This experiment targets minority Southeast Asian languages: Thai (th), Vietnamese (vi), and Indonesian (id), using GigaSpeech 2~\cite{gigaspeech2} for both training and evaluation. To mimic real-world low-resource conditions, training data is limited to 200 hours per language.
Our model uses the Whisper large-v3 encoder~\cite{radford2023robust}, ReLU-activated MLP layers as the projector, and Sailor-7B~\cite{sailor} as the LLM. To improve performance, we partially fine-tune the encoder by freezing its first 30 layers and updating only the final two. For the LLM, we apply LoRA adapters to the query and value projections in each self-attention layer.
Table~\ref{tab:low_resource_asr_result} reports results on low-resource ASR tasks. For Thai and Vietnamese, the LLM-based model (Whisper encoder + Sailor-7B) outperforms both Whisper and Whisper-finetune. In Indonesian, while it falls short of Whisper-finetune, it still surpasses the base Whisper model. These results suggest that incorporating LLMs effectively leverages rich textual knowledge to improve ASR under limited-resource settings.
Sailor-7B, fine-tuned for Southeast Asian languages, further enhances performance. Compared to Vicuna-based models, the Sailor-based system achieves better results across all three languages, underscoring the value of language-specific LLM selection in low-resource scenarios.

\begin{table}[htbp]
\caption{Comparison of Whisper and LLM-based ASR in WER/CER for low resource languages among Thai(th), Vietnamese(vi), and Indonesian(id). ``+ Fine-tuning'' refers to the model that is fine-tuned using the same amount low-resource language training data compared to the LLM-based ASR models. }
\label{tab:low_resource_asr_result}
\centering
\setlength{\tabcolsep}{5.2pt}
\renewcommand{\arraystretch}{1.1}
\begin{tabular}{lccc}
\toprule
\multirow{2}{*}{\textbf{Model}} & \multicolumn{3}{c}{\textbf{Language (WER\%) ↓}} \\
\cmidrule(lr){2-4}
& \textbf{th} & \textbf{vi} & \textbf{id} \\
\midrule
Whisper Enc. + Whisper Dec.           & 20.44 & 17.94 & 20.03 \\
\quad + Fine-tuning              & 17.08 & 16.15 & \textbf{17.13} \\
Whisper Enc. + Vicuna-7B            & 17.25 & 17.95 & 19.83 \\
Whisper Enc. + Sailor-7B         & \textbf{16.13} & \textbf{15.10} & 19.48 \\
\bottomrule
\end{tabular}
\end{table}

\paragraph{Experiments on the code switching dataset}
We also conducted experiments on the ASRU 2019 Mandarin-English Code-Switching Challenge dataset~\cite{asru_cs}, which includes 200 hours of code-switching speech and 500 hours of Mandarin-only speech. In this study, we use only the 200-hour code-switching subset for training and a 20-hour test set for evaluation.
Our system uses Whisper large-v3 as the speech encoder, Qwen2-7B as the LLM decoder, and a lightweight linear projector to align speech and text modalities. To enhance performance, LoRA adapters are applied to the query and value matrices in each self-attention layer of the LLM.
Table~\ref{tab:code_switch_result} shows that the LLM-based model outperforms Whisper large-v3 across all metrics on the code-switching test set, achieving a 20.2\% relative reduction in MER, highlighting the effectiveness of the LLM-based architecture.

\begin{table}[htbp]
\centering
\renewcommand\arraystretch{1.2}
\caption{Comparison of Whisper large-v3 and LLM-based ASR in Mandarin-English code switching ASR. MER denotes mixed error rate for both Chinese character and English words.}
\label{tab:code_switch_result}
\tabcolsep=0.3cm
\begin{tabular}{c|ccc}
\hline
\textbf{Model}        & \textbf{CN CER} & \textbf{EN WER} & \textbf{MER}  \\ 
\hline
Whisper large-v3               & 7.85            & 35.18           & 10.01         \\ 
Ours   & \textbf{5.97}            & \textbf{26.84}           & \textbf{7.99}          \\ 
\hline
\end{tabular}
\end{table}

\subsection{Contextual Automatic Speech Recognition (CASR)}

\subsubsection{Task setup}

\begin{figure*}[htbp]
\centering
\includegraphics[width=0.9\linewidth]{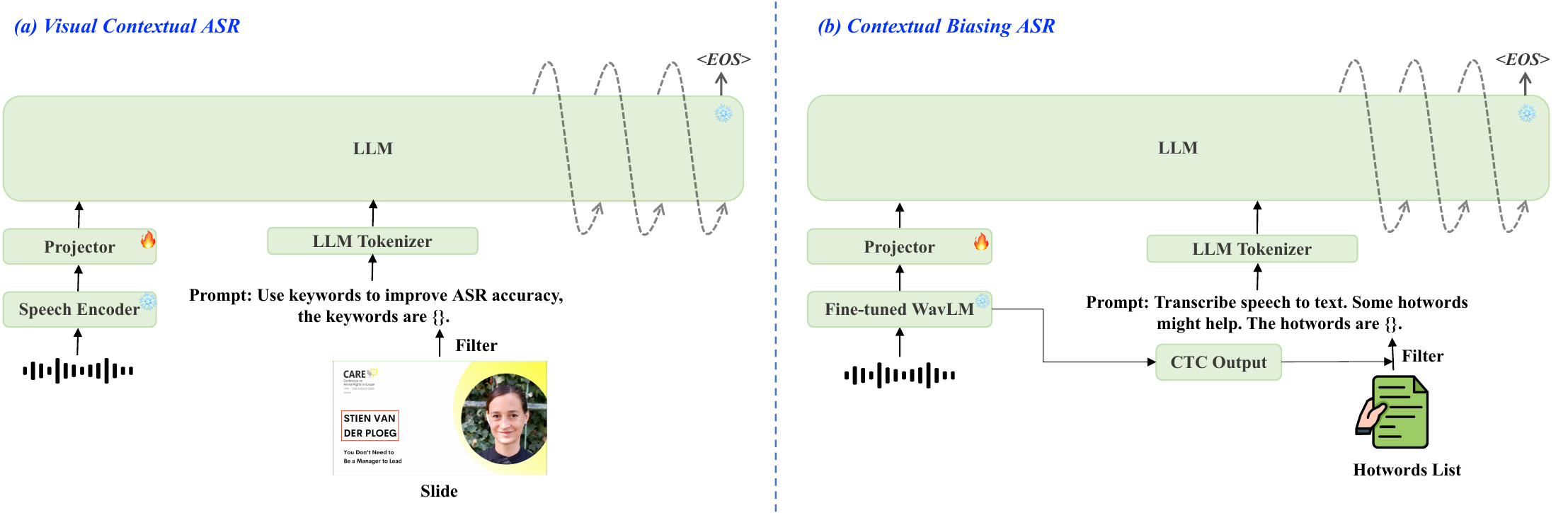}
\caption{(a) LLM-based Visual Contextual ASR in SLAM-LLM framework. (b) LLM-based Contextual Biasing ASR in SLAM-LLM framework. }
\label{fig:CASR_model}
\end{figure*}

We investigate two types of Contextual ASR tasks: \textbf{Visual Contextual Speech Recognition} and \textbf{Contextual Biasing Speech Recognition}, as shown in Figure~\ref{fig:CASR_model}. 
In Visual Contextual ASR, textual keywords extracted from presentation slides are used to improve transcription accuracy for conference content. Each speech segment is paired with pre-processed OCR results and slide-specific keywords.
In Contextual Biasing ASR, a predefined list of hundreds to thousands of biasing terms—such as named entities, contact names, or song titles—is provided to help recognize rare or domain-specific vocabulary.

\subsubsection{Visual Contextual Speech Recognition}

\paragraph{Datasets}

We use the SlideSpeech~\cite{wang2023slidespeech} dataset for training and evaluation. This large-scale audio-visual corpus is constructed from YouTube conference videos and provides high-quality transcribed speech aligned with synchronized slides. It includes 720p videos, 16kHz audio, pre-processed OCR results, and extracted keywords per segment, supporting multimodal ASR tasks.
Two training sets are available: a large set (L95) with 473 hours and a smaller subset (S95) with 161 hours. The development and test sets contain 5.07 and 8.75 hours, respectively. The dataset’s alignment between speech and slides makes it well-suited for text-enhanced multimodal ASR, particularly in correcting domain-specific or proprietary terms often misrecognized by conventional systems.

\paragraph{Model setup}

We adopt the official WavLM Large model as the speech encoder, Vicuna-7B as the LLM decoder, and a lightweight linear projector to align speech features with the LLM input space. The projector first downsamples 50Hz features to 10Hz via a 1D convolution, followed by two linear layers with a hidden dimension of 2048. The training recipes and model checkpoints can be found here\footnote{\url{https://github.com/X-LANCE/SLAM-LLM/tree/main/examples/mala_asr_slidespeech}}.

\paragraph{Experiments}

Table~\ref{tab:keyword} reports performance on the SlideSpeech dataset using WER, biased WER (B-WER), unbiased WER (U-WER), and keyword Recall.
Our baseline model, trained on L95/S95, achieves WERs of 9.4\%/11.7\%, showing relative reductions of 27.9\%/44.7\% over the SlideSpeech contextual ASR baseline. Incorporating keyword prompts further reduces WERs to 9.0\%/11.2\%, improving by 3.6\%/4.1\%. Notably, B-WER drops from 10.8\% to 5.8\% on L95 and from 14.9\% to 8.3\% on S95, while Recall improves from 89.4\% to 94.4\% and from 85.3\% to 92.0\%. U-WER remains stable, indicating effective use of keyword prompts without harming general transcription.
Fine-tuning the LLM with LoRA adapters further improves results. On L95, WER decreases from 9.4\% to 8.7\%, and to 8.4\% when keyword prompts are used.
Compared to prior contextual ASR systems, including CPP~\cite{wang2023slidespeech}, LCB-Net~\cite{yu2023lcbnet}, and the SlideSpeech baseline, our LLM-based approach achieves significantly lower WER, highlighting its superior capability in leveraging context for accurate transcription.

\begin{table*}[htbp]
  \caption{ASR results of with or without contextual keywords evaluated on dev/test datasets, trained on S95/L95 datasets, compared with previous NN-based models.}
  \label{tab:keyword}
  \centering
  \begin{tabular}{llccccccccc}
    \toprule
    \multirow{2}{*}{\textbf{Train}} & 
    \multirow{2}{*}{\textbf{Model}} & 
    \multirow{2}{*}{\makecell{ \textbf{Contextual}\\ \textbf{Keywords}}} &
    \multicolumn{4}{c}{\textbf{Dev}} & \multicolumn{4}{c}{\textbf{Test}} \\
    \cmidrule(lr){4-7} \cmidrule(lr){8-11}
    & & & \textbf{WER} & \textbf{B-WER} & \textbf{U-WER} & \textbf{Recall $\uparrow$} & \textbf{WER} & \textbf{B-WER} & \textbf{U-WER} & \textbf{Recall $\uparrow$}  \\
    \midrule
    \multirow{5}{*}{\makecell{S95\\(161h)}} & 
    SlidesSpeech~\cite{wang2023slidespeech} & \ding{55}& 21.05 & 31.27 & 20.29 & 68.76 & 21.22 & 26.60 & 20.83 & 73.51 \\
     & CPP~\cite{wang2023slidespeech} &\ding{51}& 20.80 & 28.61 & 20.22 & 71.48 & 20.95 & 24.05 & 20.73 & 76.10 \\
     & LCB-net~\cite{yu2023lcbnet} &\ding{51}& 18.80 & 27.90 & 18.11 & 72.09 & 19.21 & 23.70 & 18.89 & 76.48 \\
     & Ours & \ding{55}& 11.57 & 16.23 & \textbf{11.28} & 83.83 & 11.80 & 13.52 & 11.71 & 86.71 \\ 
     & Ours &\ding{51}&  \textbf{11.14} & \textbf{8.92} & 11.36 & \textbf{91.44} & \textbf{11.26} & \textbf{7.67} & \textbf{11.52} & \textbf{92.50} \\
    \midrule
    \multirow{7}{*}{\makecell{L95\\(473h)}} & SlidesSpeech~\cite{wang2023slidespeech} & \ding{55}& 13.09 & 16.13 & 12.87 & 83.90 & 12.89 & 12.70 & 12.90 & 87.43 \\
     & CPP~\cite{wang2023slidespeech} & \ding{51}& 12.64 & 12.39 & 12.66 & 87.64 & 12.38 & 9.32 & 12.60 & 90.86\\ 
     & LCB-net~\cite{yu2023lcbnet} &\ding{51}& 12.21 & 12.12 & 12.21 & 87.98 & 12.02 & 9.03 & 12.24 & 91.12 \\
     & Ours & \ding{55}&   9.38 & 11.98 & 9.19 & 88.08 & 9.34& 9.52 & 9.33 & 90.64 \\ 
     & \quad + LoRA & \ding{55}&  8.82 & 9.62 & 8.77 & 90.38 & 8.61& 7.34 & \textbf{8.72} & 92.84 \\
     & Ours & \ding{51}& 8.91 & 6.07 & 9.13 & 94.02 & 9.14 & 5.47 & 9.42 & 94.87 \\
     & \quad + LoRA & \ding{51}& \textbf{8.30} & \textbf{5.22} & \textbf{8.53} & \textbf{94.87} & \textbf{8.46} & \textbf{4.89} & 8.73 & \textbf{95.31} \\
    \bottomrule
  \end{tabular}
\end{table*}

\subsubsection{Contextual Biasing Speech Recognition}

\paragraph{Datasets}
We use the LibriSpeech corpus for training and evaluation, following prior work~\cite{le2021contextualized,huang2023contextualized}. The LLM-based ASR model is trained on the full 960-hour training set using the official WavLM Large as the pre-trained encoder. Evaluation is conducted on the standard dev-clean/dev-other and test-clean/test-other sets.
For contextual ASR, we adopt the artificial biasing list from~\cite{le2021contextualized}, where the 5,000 most frequent training words are labeled as common, and the rest as rare. Each test-time biasing list includes rare words from the reference and distractors sampled from the 209.2K rare-word vocabulary. Lists of size $N={100, 500, 1000, 2000}$ are constructed by varying the number of distractors.

\paragraph{Model setup}

We fully fine-tune the official WavLM Large model (315.5M parameters) on the 960-hour LibriSpeech training set using CTC loss. The fine-tuned model serves as the speech encoder. The rest of the architecture mirrors the visual contextual ASR setup, employing Vicuna-7B (6.7B) as the LLM decoder and a lightweight linear projector (15.7M). The training recipes and model checkpoints can be found here\footnote{\url{https://github.com/X-LANCE/SLAM-LLM/tree/main/examples/contextual_asr}}.

\paragraph{Experiments with different biasing lists}

Table~\ref{tab:wer} presents the performance of our CTC-assisted LLM-based contextual ASR (CASR) model on LibriSpeech, evaluated using WER, B-WER, and U-WER.
For non-contextual ASR, using the pre-trained WavLM Large encoder yields WERs of 2.13\%/4.73\% on test-clean/test-other. Fine-tuning slightly improves performance to 2.11\%/4.20\%, serving as our baseline.
In contextual ASR, prompting with an empty string achieves WER/B-WER of 1.96\%/9.33\% (test-clean) and 4.18\%/20.02\% (test-other), showing robustness even without explicit hotwords. Supplying ground-truth hotwords further reduces WER/B-WER to 1.13\%/2.78\% and 2.68\%/6.00\%, representing the upper-bound performance.
In practical settings, where biasing lists mix relevant terms with distractors, best performance is achieved with 100-word lists, reducing WER/B-WER to 1.27\%/3.67\% and 2.72\%/8.02\%, corresponding to relative WER/B-WER reductions of 39.81\%/63.37\% and 35.24\%/61.37\% over the baseline, while U-WER remains stable.
As list size increases, performance degrades slightly, but even with 2,000-word lists, the model achieves 1.38\%/4.41\% (test-clean) and 3.20\%/10.02\% (test-other), maintaining notable improvements. These results demonstrate the model’s strong capacity to filter and utilize biasing information effectively.

\begin{table}[ht]
\centering
\setlength\tabcolsep{1pt}
\scriptsize
\caption{Performance of LLM-Based Contextual ASR on LibriSpeech test sets. ``No bias" indicates adding an empty string, ``Bias List" refers to incorporating filtered hotwords from the complete biasing list, and ``GT Hotwords" denotes including the exact ground truth hotwords during inference.}
\resizebox{\linewidth}{!}{
\begin{tabular}{lcccccccc}
\toprule
\multirow{2}{*}{\textbf{WavLM Encoder}} & \multirow{2}{*}{\textbf{Prompt}} & \multicolumn{3}{c}{\textbf{test-clean}} & \multicolumn{3}{c}{\textbf{test-other}} \\
\cmidrule(lr){3-5} \cmidrule(lr){6-8}
& & \textbf{WER} & \textbf{B-WER} & \textbf{U-WER} & \textbf{WER} & \textbf{B-WER} & \textbf{U-WER} \\
\midrule
Pre-trained & \ding{55} & 2.13 & 10.15 & 1.20 & 4.73 & 22.43 & 2.84 \\
CTC Fine-tuned & \ding{55} & 2.11 & 10.02 & 1.20 & 4.20 & 20.76 & 2.43 \\
\midrule
\multirow{7}{*}{CTC Fine-tuned} & No bias & 1.96 & 9.33 & 1.11 & 4.18 & 20.02 & 2.49 \\
\cmidrule(lr){2-8}
& 100 $_{\text{Biasing words}}$ & 1.27 & 3.67 & 1.00 & 2.72 & 8.02 & 2.16 \\
& 500 $_{\text{Biasing words}}$ & 1.33 & 3.92 & 1.03 & 3.04 & 9.04 & 2.40 \\
& 1000 $_{\text{Biasing words}}$ & 1.33 & 4.16 & 1.00 & 2.99 & 9.33 & 2.31 \\
& 2000 $_{\text{Biasing words}}$ & 1.38 & 4.41 & 1.03 & 3.20 & 10.02 & 2.47 \\
\cmidrule(lr){2-8}
& GT Hotwords & \textbf{1.13} & \textbf{2.78} & \textbf{0.94} & \textbf{2.68} & \textbf{6.00} & \textbf{2.32} \\
\bottomrule
\end{tabular}
}
\label{tab:wer}
\end{table}

\begin{table}[htbp]
\setlength\tabcolsep{1.5pt}
\centering
\caption{Performance comparison with previous NN-based contextual ASR models utilizing artificial biasing lists proposed in~\cite{le2021contextualized} on the Librispeech test sets, with biasing list size 
N set to 1000.}
\begin{tabular}{lcccccc}
\toprule
\multirow{2}{*}{\textbf{Model}} & \multicolumn{3}{c}{\textbf{test-clean}} & \multicolumn{3}{c}{\textbf{test-other}} \\ \cmidrule(lr){2-4} \cmidrule(lr){5-7}
 & \textbf{WER} & \textbf{B-WER} & \textbf{U-WER} & \textbf{WER}& \textbf{B-WER} & \textbf{U-WER} \\
\midrule
CPPNet~\cite{huang2023contextualized} & 3.81 & 11.40 & 2.90 & 8.75 & 25.30 & 6.90 \\ 
Deep Biasing+BPB~\cite{sudo2024contextualized}& 3.47 & 7.70 & 3.00 & 7.34 & 15.80 & 6.40 \\ 
TCPGen$_{\text{+GNN enc.}}$ ~\cite{sun2022tree}& 3.10 & 6.70 & - & 7.90 & 17.80 & - \\ 
GA-CTC ~\cite{tang2024improving}& 2.40 & 6.30& 2.00 & 6.20 & 15.20 & 5.20 \\ 
TCPGen$_{\text{+p+phn-awareQ}}$\cite{futami2024phoneme}& 2.20 & 4.60 & - & 6.00 & 12.30 & - \\ 
DB-NNLM ~\cite{le2021contextualized}& 2.14 & 6.70 & 1.60 & 6.35 & 17.20 & 5.10 \\ \hline 
\\[-2ex]
\textbf{Ours} ~\cite{yang2024ctc}& \textbf{1.33} & \textbf{4.16} & \textbf{1.00} & \textbf{2.99} & \textbf{9.33} & \textbf{2.31} \\ 
\bottomrule
\end{tabular}
\label{tab:compare}
\end{table}

\paragraph{Comparison with Non-LLM-based CASR models}
Table \ref{tab:compare} shows a performance comparison with various NN-based contextual ASR models using the artificial biasing lists proposed in ~\cite{le2021contextualized} on the Librispeech test sets, with the biasing list size set to 1000. Our LLM-based method significantly outperforms traditional NN-based approaches.

\begin{figure}[htbp]
\centering
\includegraphics[width=0.8\linewidth]{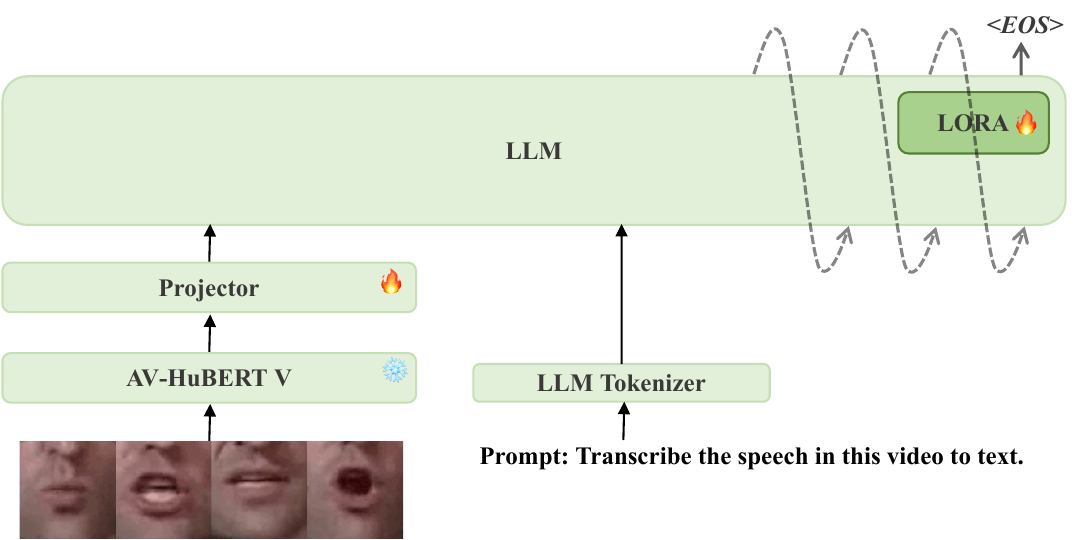}
\caption{The model architecture of LLM-based visual speech recognition (VSR) in SLAM-LLM framework. }
\label{fig:vsr_model}
\end{figure}

\subsection{Visual Speech Recognition (VSR)}

Visual Speech Recognition (VSR) aims to transcribe speech by analyzing visual cues from a speaker's face, as shown in Figure~\ref{fig:vsr_model}. We explore LLM-based VSR in our SLAM-LLM framework.

\paragraph{Datasets}
We conduct experiments on the large-scale AVSR benchmark dataset LRS3~\cite{afouras2018lrs3}, which comprises over 400 hours of TED and TEDx videos with aligned subtitles and word boundaries. The dataset is split into three subsets: 119k utterances (407 hours) for pre-training, 32k utterances (30 hours) for train-val, and 1,452 utterances (1 hour) for testing. LRS3 presents a significant challenge due to its wide variation in head poses, lighting, speaking styles, and speakers.

\begin{table}[h]
\setlength\tabcolsep{3pt}
\centering
\scriptsize
\caption{Performance of LLM-based VSR model on the LRS3 dataset with SLAM-LLM framework.}
\label{tab:vsr}
\begin{tabular}{lccccc}
\toprule
\textbf{Model} & \textbf{Encoder} &  \textbf{Decoder} & \textbf{Criterion} & \makecell[c]{\textbf{Trainable}\\ \textbf{Param.}} & \textbf{WER} \\
\midrule
AV-HuBERT~\cite{shi2022learning} & \multirow{2}{*}{AV-HuBERT$_{\text{v}}$} & Transformer & CTC+Attention & 325 & 28.6 \\ 

Ours & & Vicuna$_{\text{v1.5-7B}}$ & Decoder-Only & 49 & 28.3 \\
\bottomrule
\end{tabular}
\end{table}

\paragraph{Model setup}

We use the official AV-HuBERT Large model (477.3M), pre-trained on LRS3 and VoxCeleb2~\cite{chung2018voxceleb2} (1,759 hours of unlabeled data), and fine-tuned on 433 hours of labeled LRS3 data for the VSR task. Vicuna-7B serves as the LLM decoder, with a lightweight linear projector as the adaptor.
To enhance performance, we apply LoRA adapters to the key, query, value, and output projection layers in each self-attention module of the LLM, using rank 32, alpha 32, and dropout 0.05. This introduces 33.6M additional trainable parameters. The training recipes and model checkpoints can be found here\footnote{\url{https://github.com/X-LANCE/SLAM-LLM/tree/main/examples/vsr_LRS3}}.

\paragraph{Experiments}
Table \ref{tab:vsr} shows the performance of our LLM-based VSR model trained and evaluated on the LRS3 dataset. When utilizing the same amount of labeled and unlabeled training data, our SLAM-VSR model achieves a better WER of 28.3 compared with the AV-HuBERT baseline model fine-tuned for VSR task, with much less trainable parameters of only 49 million, demonstrating the efficiency of LLM-based structure.

\subsection{Speech-to-Text Translation (S2TT)}

\paragraph{Model setup}
The LLM-based speech-to-text translation (S2TT) model in SLAM-LLM uses a frozen Whisper encoder and Qwen2-7B LLM, with the Q-Former projection as the only trainable component. The encoder extracts speech features, which are compressed and aligned to the LLM input space via the Q-Former. The LLM then generates text outputs based on concatenated auditory and textual embeddings. The training recipes and model checkpoints can be found here\footnote{\url{https://github.com/X-LANCE/SLAM-LLM/tree/main/examples/st_covost2}}.

\paragraph{Datasets}

We conduct experiments using CoVoST-2~\cite{wang2020covost} for training and MuST-C~\cite{di2019must} for zero-shot evaluation. 

\begin{table}[htbp]
  \centering
  \small
  \setlength\tabcolsep{2pt}
\caption{The prompt design is intended for instruction fine-tuning three tasks: ASR, MMT, and SRT. We designed minimalist yet effective prompts to distinguish tasks.}
\label{slam-st-pattern}
  \resizebox{1\linewidth}{!}{
    \begin{tabular}{c c c} 
    \toprule 
\textbf{Task}   &\textbf{Instruction} &\textbf{Prediction}  \\ \hline
ASR&\texttt{<|en|>}  &Will it rain tomorrow?\\  \hline
MMT&Will it rain tomorrow?\texttt{<|de|>} &Will it rain tomorrow? \texttt{<|de|>}Regnet es morgen?    \\ \midrule

SRT&\texttt{<|de|>}  &Will it rain tomorrow? \texttt{<|de|>}Regnet es morgen?    \\
\bottomrule
    \end{tabular}
    }
\end{table}

\begin{table}[htb]
  \centering
  \scriptsize 
  \setlength\tabcolsep{2pt} 
  \caption{Speech translation BLEU scores on CoVoST-2 and MuST-C datasets. We conducted experiments in German (De), Japanese (Ja), and Chinese (Zh).  We use \uline{underline} to highlight previous SOTA baseline, and use \textbf{bold} to highlight surpassing the SOTA performance.}
  \resizebox{\linewidth}{!}{ 
\begin{tabular}{lcccccccc} 
    \toprule 
    \multirow{2}{*}{En$\rightarrow$X} & \multicolumn{4}{c}{\textbf{CoVoST-2}} & \multicolumn{4}{c}{\textbf{MuST-C$_{\text{(zero-shot)}}$}} \\ 
    \cmidrule(lr){2-5} \cmidrule(lr){6-9}
    & De & Ja & Zh & Avg. & De & Ja & Zh & Avg. \\ \hline
    \textbf{Cascaded ST Methods} & & & & & & & & \\
    \hspace{1em} Whisper+NLLB$_{\text{3.3b}}$\cite{costa2022no} & - & 19.0 & 32.0 & - & - & - & 13.5 & - \\ 
    \hspace{1em} Whisper+Qwen$_{\text{7B-chat}}$\cite{wang2024blsp} & 25.3 & 22.7 & 43.6 & 30.5 & \uline{22.8} & - & - & - \\ 
    \midrule
    \textbf{End-to-End Methods} & & & & & & & & \\
    \hspace{1em} SALMONN~\cite{tang2023salmonn} & 18.6 & 22.7 & 33.1 & 24.8 & - & - & - & - \\ 
    \hspace{1em} BLSP-KD~\cite{wang2024blsp} & 24.4 & 21.3 & 41.3 & 29.0 & 20.7 & - & - & - \\ 
    \hspace{1em} SeamlessM4T-V2~\cite{barrault2023seamless} & \uline{37.0} & 23.5 & 34.6 & 31.7 & - & - & 18.1 & - \\
    \hspace{1em} Qwen2-Audio~\cite{chu2024qwen2} & 29.9 & \uline{28.6} & \uline{45.2} & \uline{34.5} & 22.4 & \uline{7.8} & 18.3 & \uline{16.1} \\  \hline
    \hspace{1em} \textbf{Ours-7B} & \textbf{28.7} & \textbf{30.8} & \textbf{47.7} & \textbf{35.7} & \textbf{18.3} & \textbf{11.3} & \textbf{21.2} & \textbf{16.9} \\
    \bottomrule
  \end{tabular}
  }
  \label{tab:slam-st-result}
\end{table}

\paragraph{Experiments}

We adopt a multimodal Chain-of-Thought (CoT) approach to decompose the speech translation (ST) task into two sequential stages: ASR followed by multimodal machine translation (MMT), collectively referred to as SRT (Speech Recognition and Translation), as illustrated in Table~\ref{slam-st-pattern}. This formulation enables end-to-end generation of both transcriptions and translations.
We apply three-stage SFT with curriculum learning on CoVoST-2, followed by zero-shot evaluation on MuST-C. As shown in Table~\ref{tab:slam-st-result}, our model achieves state-of-the-art (SOTA) performance on CoVoST-2 and outperforms existing methods in zero-shot English-to-Chinese translation on MuST-C.

\subsection{Speech Emotion Captioning (SEC)}

\paragraph{Task setup}

Speech Emotion Captioning (SEC) aims to describe speech emotions using natural language, offering a more nuanced alternative to traditional speech emotion recognition (SER), which typically relies on fixed emotion categories and struggles to capture the complexity of human affect.
SECap~\cite{xu2024secap} first introduced SEC using an Encoder–Projector–LLM framework to generate rich emotional descriptions. SEC has enabled applications such as PerceptiveAgent~\cite{yan2024talk}, which enhances emotional awareness in conversations, and AVI-Talking~\cite{sun2024avi}, which improves the expressiveness of talking-face animations. Beyond these, SEC holds promise for a wide range of yet-unexplored applications.

\paragraph{Datasets}
We trained our model using approximately $40$ hours of in-house emotional speech data. 
Each audio segment in the dataset is annotated with $1$ to $3$ emotion-related captions, which provide natural language descriptions of the expressed affective content. 
These captions were carefully curated to capture a diverse range of emotional nuances beyond traditional categorical labels, enabling the model to learn fine-grained emotional representations. 

\paragraph{Model Setup}

SECap~\cite{xu2024secap} achieves state-of-the-art SEC performance using HuBERT~\cite{hsu2021hubert} as the audio encoder, a Chinese LLaMA-2 decoder, and a Q-Former Bridge-Net to extract emotion-relevant acoustic features. However, it relies on complex training strategies—Speech–Transcription Mutual Information Learning (STMIL) and Speech–Caption Contrastive Learning (SCCL)—to attain its reported gains.
To enhance intrinsic emotion perception while simplifying the training pipeline, we replace the audio encoder with emotion2vec~\cite{ma2023emotion2vec}, a self-supervised model tailored for emotion representation. We adopt Vicuna-7B as the LLM and retain the Q-Former as the projection module. The training recipes and model checkpoints can be found here\footnote{\url{https://github.com/X-LANCE/SLAM-LLM/tree/main/examples/sec_emotioncaps}}.

\paragraph{Experiments}
Following the evaluation protocol in SECap, we test the model on the 600 sentences from the EMOSpeech testset released by~\cite{xu2024secap}, and we report sentence similarity (SIM) based on MACBERT~\cite{cui2021pre} as the primary metric.
Table \ref{tab:sec} contrasts our LLM-based SEC model in SLAM-LLM with the SECap family and the HTSAT–BART baseline reported in the SECap paper. 
Without any auxiliary objectives, our model attains a SIM of 71.10\%, outperforming the vanilla SECap and narrowing the gap to the most heavily engineered SECap variant. Relative to the HTSAT–BART baseline, the gain exceeds 7 percentage points, underscoring the effectiveness of coupling the SSL emotion2vec encoder with a strong LLM decoder. 

\begin{table}[htbp]
\centering
\caption{Performance of speech emotion captioning with SLAM-LLM framework. The specific information of the different modules is given in the table. Results other than ours come from the SECap paper.}
\label{tab:sec}
 \resizebox{\linewidth}{!}{
\begin{tabular}{lccc}
\toprule
\textbf{Model} & \textbf{Encoder} &  \textbf{Decoder} &  \textbf{SIM(\%)} \\
\midrule
Baseline~\cite{xu2024secap} & HTSAT & BART & 63.95 \\
\hline
SECap~\cite{xu2024secap} &  \multirow{3}{*}{HuBERT} &  \multirow{3}{*}{Chinese-LLaMA2-7B} & 67.29 \\
\quad+STMIL & & & 68.75 \\
\quad\quad+SCCL & & & 71.95 \\
\hline
Ours & emotion2vec & Vicuna-v1.5-7B & 71.10 \\
\bottomrule
\end{tabular}
}
\end{table}

\subsection{Takeaways}
From the above experiments, several key takeaways can help guide future research and development of LLM-based speech tasks:
\begin{itemize}
    \item \textbf{Larger models improve performance.} Obviously, both larger speech encoders and larger LLMs consistently lead to better ASR performance.
    \item \textbf{Chat LLMs outperform pre-trained LLMs.} Chat models, such as Vicuna-7B, consistently outperform pre-trained LLMs in ASR tasks, particularly when paired with larger speech encoders.
    \item \textbf{Self-supervised encoders are superior at scale.} Self-supervised encoders outperform supervised encoders once they reach a sufficient size. Fine-tuning these self-supervised encoders yields significant performance gains, and though our experiments were limited, we believe fine-tuning LLMs on in-domain text data would also enhance performance. 
    \item \textbf{Domain specialization LLMs are good helpers.} For instance, utilizing LLMs trained on low-resource languages can enhance the effectiveness of LLM-based low-resource languages ASR. 
    \item \textbf{Whisper models face limitations with truncation.} Whisper models, which pad speech inputs to 30 seconds, experience performance degradation when truncated. Utterance with 30s also significantly increases computational demands during LLM post-training. We recommend using self-supervised models with variable-length embeddings as encoders to mitigate this issue. 
\end{itemize}
More detailed and technique-specific information can be found in sub-tasks research~\cite{ma2024embarrassingly, yang2024mala, yang2024ctc} of the SLAM-LLM series. 
\section{LLM-based Audio and Music Processing}

\subsection{Automated audio captioning (AAC)}
\begin{figure*}[htbp]
\centering
\includegraphics[width=0.9\linewidth]{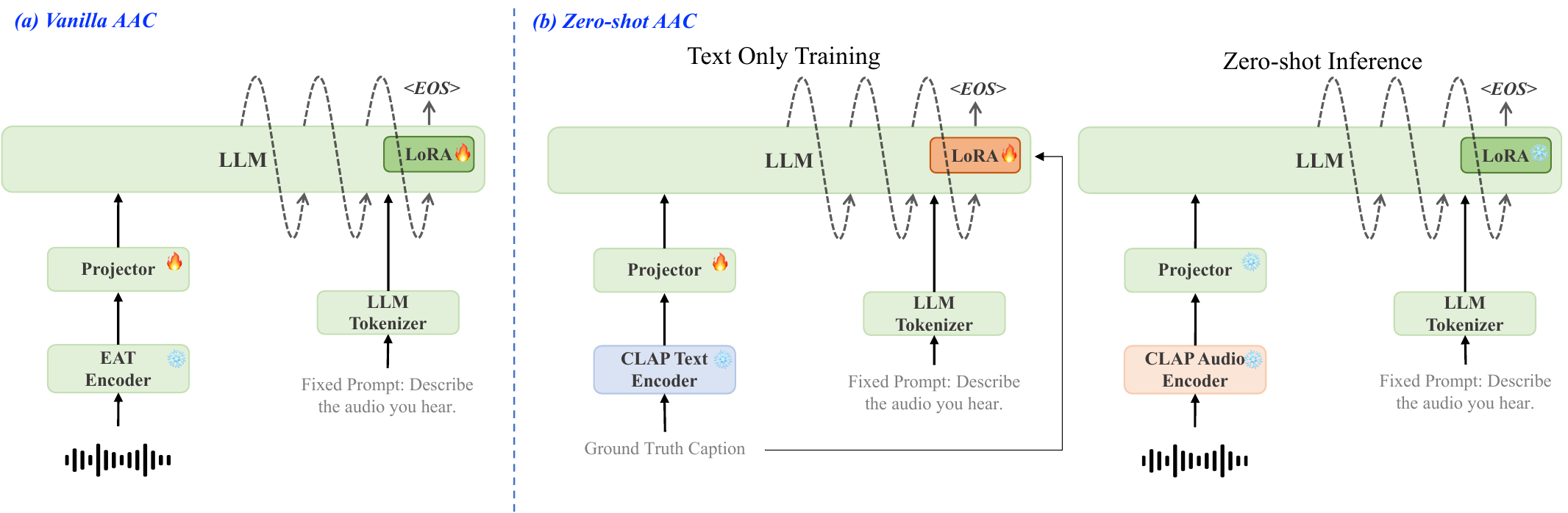}
\caption{(a). LLM-based vanilla automated audio captioning in SLAM-LLM framework (b). LLM-based zero-shot automated audio captioning in SLAM-LLM framework}
\label{fig:slam-aac-all}
\end{figure*}

Automated audio captioning (AAC) aims to generate fine-grained natural language descriptions from input audio, serving as a key task in audio processing. 
As shown in Figure~\ref{fig:slam-aac-all}, we focus on two mainstream AAC paradigms with the SLAM-LLM framework: (1) \textbf{The vanilla setting}, also known as \textit{fully supervised} audio captioning, where the model is trained on paired audio-text data. (2) \textbf{The zero-shot setting}, where the model is trained exclusively on text data and is expected to generate captions for audio clips in a zero-shot manner during inference.

\paragraph{Datasets} 
\label{para:AAC_datasets}
We use four key audio-text datasets for the audio captioning experiments: Clotho~\cite{drossos2020clotho}, AudioCaps~\cite{kim2019audiocaps}, WavCaps~\cite{mei2023wavcaps}, and MACS~\cite{martin2021ground}. For Clotho, version 2.1 was used, consisting of audio clips ranging from 15 to 30 seconds in duration. The dataset includes 3,839 training examples, 1,045 validation examples, and 1,045 evaluation examples, with each audio clip accompanied by five captions. AudioCaps contains over 50,000 ten-second audio clips derived from AudioSet~\cite{gemmeke2017audio}. It is split into a training set (49,274 clips, each with one caption), a validation set (494 clips, each with five captions), and a test set (957 clips, each with five captions). WavCaps comprises 403,050 audio clips collected from multiple sources, including AudioSet, BBC Sound Effects, FreeSound, and SoundBible. MACS consists of 3,930 ten-second audio files, each associated with 2 to 5 captions, mainly recorded in three acoustic environments (airport, public square, and park). For our experiments, we used the training sets from Clotho, AudioCaps, and MACS, along with the entire WavCaps dataset for pre-training. 
In addition, the Clotho training set was augmented using a paraphrasing method, which draws from the back-translation \cite{sennrich2015improving} technique used in machine translation to expand the dataset during pre-training.

\paragraph{Evaluation Metrics} 
To evaluate the quality of generated audio captions, we used several standard AAC metrics: METEOR \cite{banerjee2005meteor}, CIDEr \cite{vedantam2015cider}, SPICE \cite{anderson2016spice}, SPIDEr \cite{liu2017improved}, SPIDEr-FL \cite{zhou2022can} and FENSE \cite{zhou2022can}. METEOR considers unigram precision, recall, synonyms, and stemming. CIDEr measures n-gram consensus using TF-IDF. SPICE compares semantic graphs of generated and reference captions. SPIDEr linearly combines CIDEr and SPICE for balanced evaluation. SPIDEr-FL further incorporates fluency detection from FENSE, which uses Sentence-BERT for semantic similarity combined with fluency error detection.

\paragraph{Model Setup} 
For vanilla setting, we utilize the frozen EAT model~\cite{chen2024eat} as the audio encoder to extract fine-grained audio representations, which are then downsampled and aligned with LLM embeddings through a linear projector. Specifically, the projector downsamples the 50Hz features to 10Hz using two linear layers, with an intermediate hidden layer dimension of 2048. For decoding, the LLM, Vicuna \cite{chiang2023vicuna}, generates captions based on these concatenated representations and is efficiently fine-tuned using LoRA. During inference, multiple candidate captions are generated through beam search within different beam widths, with the most audio-aligned caption selected as the final output using the CLAP-Refine \cite{chen2024slam} strategy. 
The training recipes and model checkpoints can be found here\footnote{\url{https://github.com/X-LANCE/SLAM-LLM/tree/main/examples/slam_aac}}. 

For zero-shot audio captioning, we utilize the frozen CLAP \cite{wu2023large} model as the audio \& text encoder. 
During training, raw captions are first encoded by the text branch of the CLAP model, and then get aligned with the LLM through a two-layer linear projector. The LLM, efficiently fine-tuned using the LoRA \cite{hu2021lora} method, learns to re-construct the ground-truth caption conditioned on the mapped CLAP latent and an encoded prompt retrieved from a datastore. 
During inference, the text branch is replaced by the audio branch of the CLAP model, and the system describes audio clips in a zero-shot manner.
Projection-based decoding~\cite{li2023decap} and retrieval-augmented generation (RAG) are employed to reduce the modality gap and improve caption quality. Audio embeddings are projected onto the text embedding space, while similar captions retrieved from a datastore are used as prompts to guide the LLM.
The training recipes and model checkpoints can be found here \footnote{\url{https://github.com/X-LANCE/SLAM-LLM/tree/main/examples/drcap_zeroshot_aac}}. 

\paragraph{Vanilla Audio Captioning Experiments}

For the vanilla audio captioning system, we trained our model on the pre-training dataset, followed by fine-tuning on the AudioCaps and Clotho datasets individually. Furthermore, we conducted a comprehensive ablation study on these two datasets to evaluate the impact of each component. 

Table \ref{tab:aac_main} compares the performance of our model with previous SOTA AAC systems. We include three models using non-LLM-based decoding: EnCLAP \cite{kim2024enclap}, WavCaps \cite{mei2023wavcaps}, and Wu et al. \cite{wu2023beats}—and two employing LLM-based decoding—Tang et al. \cite{tang2024extending} and LOAE \cite{liu2024enhancing}.
Overall, LLM-based models tend to produce higher-quality captions, likely due to their stronger reasoning capabilities. Compared to previous models, our proposed system demonstrates consistent improvements across both the Clotho and AudioCaps datasets. On Clotho, it achieves the highest scores across all AAC metrics, with a notable improvement in FENSE (54.0\%), surpassing both the LOAE and Wu et al. models. On AudioCaps, the improvements are even more pronounced: our model attains a CIDEr score of 84.1\%, significantly exceeding LOAE (81.6\%), and achieves a FENSE score of 66.8\%, outperforming all other models.

Table \ref{table:aac_ablation} shows the results of a comprehensive ablation study conducted on the AudioCaps and Clotho. The study explored the impact of different audio encoders, versions of audio encoders (pre-trained vs. fine-tuned on AudioSet), whether pre-training was performed, the use of parameter-efficient fine-tuning (e.g., LoRA), and different decoding strategies (beam search vs. CLAP-Refine). 
Results show that fine-tuned audio encoders, particularly those with strong performance in classification tasks, significantly enhance AAC quality. Pre-training the LLM-based AAC model further boosts performance, while LoRA improves training efficiency. Additionally, the CLAP-Refine decoding method consistently improves caption quality by enhancing alignment with input audio, leading to higher CIDEr and SPIDEr scores.

\begin{table*}[htbp]
\centering
\caption{
Performance comparison of  Vanilla \textit{(fully supervised)} AAC models on Clotho and AudioCaps evaluation split. 
The pre-training datasets used include AudioCaps (AC), Clotho (CL), WavCaps (WC), MACS (MA), LibriSpeech (LS)~\cite{panayotov2015librispeech}, and GigaSpeech (GS)~\cite{chen2021gigaspeech}. 
Metrics reported are METEOR (MT), CIDEr (CD), SPICE (SC), SPIDEr (SD), SPIDEr-FL (SF), and FENSE (FS). 
CL$_{C}$ and CL$_{P}$ denote the Clotho training set augmented with the ChatGPT Mix-up method and the paraphrasing approach, respectively. All metrics are reported with higher values indicating better performance. 
}
\begin{tabular}{lccccccccccccccc}
\toprule
\multirow{2.5}{*}{\textbf{Model}} & \multirow{2.5}{*}{\textbf{Pre-training Data}}  & \multicolumn{6}{c}{\textbf{Clotho Evaluation (\%)} } & & \multicolumn{6}{c}{\textbf{AudioCaps Evaluation (\%)}} \\
\cmidrule(lr){3-8} \cmidrule(lr){10-15}
 &  &  \textbf{MT} & \textbf{CD} & \textbf{SC} & \textbf{SD} & \textbf{SF} & \textbf{FS} && \textbf{MT} & \textbf{CD} & \textbf{SC} & \textbf{SD} & \textbf{SF} & \textbf{FS}  \\
\midrule
EnCLAP-large \cite{kim2024enclap} & AC+CL & 18.6 & 46.4 & 13.3 & 29.9 & ~28.9$^\ast$  & ~50.7$^\ast$ && 25.5 & 80.3 & 18.8 & 49.5 & ~49.9$^\ast$ & ~65.5$^\ast$
\\
 WavCaps$^\dag$ \cite{mei2023wavcaps} & AC+CL+WC & 18.5 & 48.8 & 13.3 & 31.0 &  ~29.6$^\ast$ & ~50.1$^\ast$ && 25.0 & 78.7 & 18.2 & 48.5 & ~48.3$^\ast$ & ~64.2$^\ast$
 \\
Wu et al. \cite{wu2023beats} & AC+CL$_{C}$ & 19.3 &  50.6  &  14.6  & 32.6 & 32.6 &  53.6  &&  -  & -  &  -  & - & - & -  
\\
Tang et al. \cite{tang2024extending} & AC+CL+WC+LS+GS & - &  -  &  -  & 31.8 & - &  -  &&  -  & -  &  -  & 50.6 & - & -  
\\
LOAE \cite{liu2024enhancing} & AC+CL+WC & 19.7 & 51.4 & 14.7 & 33.1 & 32.8 & 53.2  &&  26.7 & 81.6 & 19.3 & 50.5 & 50.4 & 66.2
\\
\hline
\\[-2ex] 
Ours & AC+CL$_{P}$+WC+MA & \textbf{19.7} & \textbf{51.5} & \textbf{14.8} & \textbf{33.2} &  \textbf{33.0} & \textbf{54.0} && \textbf{26.8} & \textbf{84.1} & \textbf{19.4} & \textbf{51.8} & \textbf{51.5} &\textbf{66.8}   \\
\bottomrule
\multicolumn{16}{l}{$^\ast$For open-source models, we evaluated metrics not reported in the original papers using our evaluation split.} \\
\multicolumn{16}{l}{$^\dag$The best performance of WavCaps on both datasets is selected for comparison (model architectures may differ).} 
\end{tabular}
\label{tab:aac_main}
\end{table*}

\begin{table}[htbp]
\centering
\setlength\tabcolsep{1pt}
\caption{Ablation study on AudioCaps and Clotho. \textit{PT} indicates a pre-trained model, \textit{FT} denotes a fine-tuned version. BS denotes beam search decoding, CR represents CLAP-Refine.}
\scriptsize
\resizebox{1\linewidth}{!}{
\begin{tabular}{lccccccccc}
\toprule
\multirow{2.5}{*}{\textbf{Dataset}} & \multirow{1.5}{*}{\textbf{Audio}}  & \multirow{1.5}{*}{\textbf{Pre-Trained}} & \multirow{1.5}{*}{\textbf{PEFT}} & \multirow{2.5}{*}{\textbf{Decoding}} & \multicolumn{4}{c}{\textbf{Metrics (\%)}} \\
\cmidrule(lr){6-9}
 & \textbf{Encoder}  & \textbf{Model} & \textbf{(LoRA)} &  & \textbf{MT} & \textbf{CD} & \textbf{SC} & \textbf{SD} \\
\midrule
\multirow{8}{*}{AudioCaps} 
 & $\text{BEATs}_{\textit{PT}}$ &   \usym{2717} & \usym{2717} & BS & 23.2 & 69.6 & 17.2 & 43.4 \\
 & $\text{BEATs}_{\textit{FT}}$ &   \usym{2717} & \usym{2717} & BS & 23.9 & 69.8 & 17.3 & 43.5 \\
 &  $\text{EAT}_{\textit{PT}}$ &    \usym{2717} & \usym{2717} & BS & 23.5 & 67.0 & 17.3 & 42.2 \\
 & $\text{EAT}_{\textit{FT}}$ &  \usym{2717} & \usym{2717} & BS & 25.0 & 75.3 & 18.5 & 46.9 \\
 & $\text{EAT}_{\textit{FT}}$ &   \usym{2717} & \usym{2713} & BS & 26.0 & 80.0 & 18.3 & 49.2 \\
 & $\text{EAT}_{\textit{FT}}$ &   \usym{2713} & \usym{2713} & BS &  26.1 & 82.7 & \textbf{19.5} & 51.1  \\
  & $\text{EAT}_{\textit{FT}}$ &   \usym{2717} & \usym{2713} & CR &  26.6 & 81.4 & 19.4 & 50.4   \\
 & $\text{EAT}_{\textit{FT}}$ &  \usym{2713} & \usym{2713} & CR &  \textbf{26.8} & \textbf{84.1} & 19.4 & \textbf{51.8}  \\
\midrule
\multirow{6}{*}{Clotho} 
 & $\text{EAT}_{\textit{FT}}$ &   \usym{2717} & \usym{2717} & BS & 17.1 & 34.1 & 11.3 & 22.7 \\
 & $\text{EAT}_{\textit{FT}}$ &   \usym{2717} & \usym{2713} & BS & 17.0 & 38.9 & 12.0 & 25.4 \\
 & $\text{EAT}_{\textit{FT}}$ &   \usym{2713} & \usym{2713} & BS & 19.2 & 49.8 & 14.4 & 32.1 \\
  & $\text{EAT}_{\textit{FT}}$ &   \usym{2717} & \usym{2713} & BS &  19.4 & 51.0 & 14.6 & 32.8  \\
  & $\text{EAT}_{\textit{FT}}$ &   \usym{2713} & \usym{2713} & CR & \textbf{19.7} & \textbf{51.5} & \textbf{14.8} & \textbf{33.2}   \\
\bottomrule
\end{tabular}
}
\label{table:aac_ablation}
\end{table}

\begin{table}[htbp]
\centering
\setlength\tabcolsep{1pt}
\caption{Performance comparison of zero-shot audio captioning systems. For metrics, MT: METEOR, CD: CIDEr, SP: SPICE, SD: SPIDEr, FS: FENSE. Higher values indicate better performance for all metrics. }
\footnotesize
\resizebox{1\linewidth}{!}{
\begin{tabular}{lcccccccccccc}
\toprule
\multirow{3}{*}{\textbf{Method}} & \multirow{3}{*}{\textbf{Scenario}}  & \multicolumn{5}{c}{\textbf{Clotho Evaluation} (\%)} & & \multicolumn{5}{c}{\textbf{AudioCaps Evaluation} (\%)} \\
\cmidrule(lr){3-7} \cmidrule(lr){9-13}
  &  &  \textbf{MT} & \textbf{CD} & \textbf{SP} & \textbf{SD} & \textbf{FS} && \textbf{MT} & \textbf{CD} & \textbf{SP} & \textbf{SD}  & \textbf{FS}  \\

\midrule
ZerAuCap \cite{salewski2023zero} && 9.4 & 14.0 & 5.3 & 9.7 & - && 12.3 & 28.1 & 8.6 & 18.3 & - \\
WSAC \cite{kouzelis2023weakly} &In& 17.4 & 37.1 & 12.3 & 24.7 & - && 24.1 & 63.3 & 17.3 & 40.3 & -\\
Zhang \textit{et al.} \cite{zhang2024zero} &Domain& 17.5 & 41.1 & 12.2 & 26.7 & 48.8 && 22.0 & 64.4 & 15.6 & 40.0 & - \\
Ours &  & \textbf{18.2} & \textbf{43.8} & \textbf{13.3} & \textbf{28.5} & \textbf{53.0} && \textbf{25.3} & \textbf{70.5} & \textbf{18.0} & \textbf{44.2}&\textbf{66.2} \\
\midrule 

WSAC \cite{kouzelis2023weakly} &Cross& 12.0 & 20.6 & 8.2 & 14.4 & - && 17.3 & 25.6 & 12.0 & 18.8 & -\\
Zhang \textit{et al.} \cite{zhang2024zero}&Domain& 13.2 & 24.8 & 9.3 & 17.1 & - && 18.2 & 33.7 & 12.4 & 23.0 & 52.1\\
Ours && \textbf{15.0} & \textbf{33.3} & \textbf{10.4} & \textbf{21.8} & \textbf{52.2} && \textbf{22.9} & \textbf{44.3} & \textbf{17.0} & \textbf{30.6}  &\textbf{62.6}
\\
\bottomrule
\end{tabular}
}
\label{tab:zeroshot-aac-main}
\end{table}

\paragraph{Zero-shot Audio Captioning Experiments} 
For zero-shot audio captioning, we conducted experiments in both in-domain and cross-domain scenarios to evaluate the performance.
For in-domain scenario, systems are trained and evaluated on the same dataset following the standard train \& val \& test split.
For cross-domain scenario, systems are trained on the training set of the source dataset and evaluated on the evaluation set of the target dataset. 
Specifically, for Clotho Evaluation, the source dataset is AudioCaps and for AudioCaps Evaluation, the source dataset is Clotho.

Table \ref{tab:zeroshot-aac-main} presents a performance comparison between our zero-shot AAC model and previous state-of-the-art. ZerAuCap \cite{salewski2023zero} uses CLAP to guide the LLM to generate descriptions, WSAC \cite{kouzelis2023weakly} trains a text decoder using the prefix language modeling paradigm conditioned on CLAP embeddings, while Zhang \textit{et al.} \cite{zhang2024zero} crafts soft and hard prompts to bridge the modality gap between audio and text embeddings of CLAP. As shown in Table \ref{tab:zeroshot-aac-main}, our model surpasses all competitive zero-shot audio captioning systems in in-domain scenarios by a large margin and is comparable with other fully supervised methods. For cross-domain scenarios, it achieves state-of-the-art results across all metrics, highlighting its robust domain-transfer capability.
Furthermore, we found that our model outperforms other methods in terms of the FENSE \cite{zhou2022can} score in both two scenarios. We hypothesize that this advantage is due to its ability to utilize the semantically rich joint multi-modal space of CLAP, which allows it to generate more refined captions.

Table \ref{tab:zeroshot-aac-ablation} shows a comprehensive ablation study on different core components of our zero-shot AAC model on both in-domain and cross-domain scenarios. The study explored the contribution of the retrieval-augmented generation (RAG), the use of LoRA adapter, and the effectiveness of the projection decoding (PD). 
Results show that projection-based decoding helps mitigate the modality gap, the LoRA enhances training efficiency and quality, while the retrieval-augmented generation can also boost model's performance, especially in the cross-domain scenario.

\begin{table}[t]
\centering
\scriptsize
\setlength\tabcolsep{1.5pt}
\caption{Ablation study on AudioCaps (in-domain) and AudioCaps $\Rightarrow$ Clotho (cross-domain) for our zero-shot AAC model DRCap. PD donates Projection Decoding. }
\resizebox{\linewidth}{!}{
\begin{tabular}{lccccc|ccccc}
\toprule
\multirow{2.5}{*}{\textbf{Main Components}} 
& \multicolumn{5}{c|}{\textbf{AudioCaps (\%)}} 
& \multicolumn{5}{c}{\textbf{AudioCaps$\Rightarrow$Clotho (\%)}} \\
\cmidrule(lr){2-6} \cmidrule(lr){7-11}
& \textbf{MT} & \textbf{CD} & \textbf{SP} & \textbf{SD} & \textbf{FS} 
& \textbf{MT} & \textbf{CD} & \textbf{SP} & \textbf{SD} & \textbf{FS} \\
\midrule
Ours & \textbf{25.5} & \textbf{70.5} & 18.0 & \textbf{44.2} & \textbf{66.2} & \textbf{15.0} & \textbf{33.3} & \textbf{10.4} & \textbf{21.8} & \textbf{52.2} \\
\quad - w/o RAG & 25.0 & 69.2 & \textbf{18.4} & 43.7 & 65.5 & 14.2 & 30.1 & 10.2 & 20.1 & 51.3 \\
\quad - w/o LoRA & 23.7 & 64.7 & 16.4 & 40.5 & 64.1 & 14.1 & 29.8 & 9.8 & 19.8 & 51.1 \\
\quad - w/o PD & 19.9 & 31.2 & 13.4 & 22.3 & 55.4 & 13.2 & 22.8 & 8.6 & 15.7 & 46.4 \\
\bottomrule
\end{tabular}
}
\label{tab:zeroshot-aac-ablation}
\end{table}

\subsection{Music Captioning (MC)}

\paragraph{Datasets} 
We utilize the LP-MusicCaps-MC datasets \cite{doh2023lp} for training, and conduct testing on the standard test set of LP-MusicCaps-MC. For all training, the music audio is sliced into 10-second clips. 
Since the raw audio of the LP-MusicCaps-MSD dataset is difficult to access, we are only able to train our models on the LP-MusicCaps-MC dataset, which is a small subset of LP-MusicCaps dataset. However, we demonstrate in the experiments that even models trained only on LP-MusicCaps-MC can achieve comparable results to existing models \cite{deng2023musilingo, doh2023lp} pre-trained on LP-MusicCaps-MSD.

\paragraph{Model setup} 
We use a pre-trained model as the music encoder and Vicuna-7b-v1.5 \cite{chiang2023vicuna} as the LLM, both of which are kept frozen during training. We use a linear projector, and the projector is the only trainable component. 
We consider two different types of music encoders. The first is the \textbf{frame-wise encoder}, which extracts features with a temporal dimension from music audio, typically trained by self-supervised learning (SSL), such as MERT \cite{li2023mert}, MusicFM \cite{won2024foundation}, and MuQ \cite{muq}. The second is the \textbf{sequence-wise encoder}, which extracts features directly from the music to extract a fixed dimensional feature (e.g., 512 or 1024 dimensions), typically trained by contrastive learning, such as Laion-CLAP \cite{wu2023large}, Microsoft-CLAP \cite{elizalde2023clap}, and MuQ-MuLan \cite{muq}.
For frame-wise encoder, the projector downsamples frame-wise features into 0.5hz, which is the final input fed to LLM, for example, 10s of music corresponds to 5 tokens. For sequence-wise encoder, we do not downsample and feed directly into the LLM after a linear layer projection, which means that the sequence-wise encoded features have only 1 token. 
The training recipes and model checkpoints can be found here\footnote{\url{https://github.com/X-LANCE/SLAM-LLM/tree/main/examples/mc_musiccaps}}.

\paragraph{Experiments} 

In Table \ref{table:slam_mc}, we compare the effect of different music encoders on the music captioning task with existing models. Note that the SLAM-LLM models in the table are all trained only on the small LP-MusicCaps-MC data. For the frame-wise encoder, MuQ is better than MusicFM and MERT, which is consistent with the basic performance of these SSL models. For the sequence-wise encoder, surprisingly, even though the sequence-wise features have only 1 token, they generally work better than the frame-wise features. Among them, MuQ-MuLan achieves the best performance in the four metrics BLEU, METEOR, ROUGE-S and BERT-S. This means that models trained using contrastive learning, even with only one embedding, are more suitable for music captioning tasks than encoders trained using SSL. We also note that although SLAM-LLM uses only a small amount of data from LP-MusicCaps-MC, it achieves comparable results with models such as MusCaps \cite{manco2021muscaps}, LP-MusicCaps \cite{doh2023lp}, and MusiLingo \cite{deng2023musilingo}, which are trained using additional pre-training data.

\begin{table}[ht]
\centering
\scriptsize
\setlength\tabcolsep{1.pt}
\caption{The results of music caption experiments, all evaluated on the LP-MusicCaps-MC test set. The rows highlighted in gray represent models pre-trained using additional or private data.}
\resizebox{\linewidth}{!}{
\begin{tabular}{ll|c|cccc}
\toprule
\multicolumn{2}{l|}{\textbf{Model}} & \textbf{Train Dataset} & \textbf{BLEU} & \textbf{METEOR} & \textbf{ROUGE-S} & \textbf{BERT-S} \\
\midrule
\rowcolor{lightgrayrow}
MusCaps & & Private & 10.2 & 17.0 & 22.2 & 83.5 \\
\rowcolor{lightgrayrow}
LP-MusicCaps &  & MSD+MTT+MC & 14.7 & 22.4 & 21.5 & 87.8 \\
\rowcolor{lightgrayrow}
MusiLingo & & MSD+MC & 30.8 & 21.6 & 21.7 & 86.8 \\
\midrule
\multirow{8}{*}{Ours} & \textit{\textbf{Encoder}}$_{\text{Frame-wise}}$ & & & & & \\
& \;MERT & \multirow{3}{*}{MC} & 22.2 & 11.6 & 16.8 & 86.3 \\
& \;MusicFM & & 25.1 & 18.6 & 19.2 & 86.2 \\
& \;MuQ & & 27.0 & 19.2 & 19.1 & 86.8 \\
\cmidrule(lr){2-7}
\multirow{4}{*}{} & \textit{\textbf{Encoder}}$_{\text{Sequence-wise}}$ & & & & & \\
& \;Laion-CLAP & \multirow{3}{*}{MC} & 25.9 & 19.6 & 20.9 & 86.6 \\
& \;Microsoft-CLAP$_\text{2023}$ & & 27.5 & 20.3 & \textbf{21.1} & 86.8 \\
& \;MuQ-MuLan & & \textbf{27.9} & \textbf{20.5} & \textbf{21.1} & \textbf{86.9} \\
\bottomrule
\end{tabular}
}
\label{table:slam_mc}
\end{table}

\subsection{Takeaways}
From the above experiments, several key takeaways can help guide future research and development of LLM-based audio and music tasks:
\begin{itemize}
    \item \textbf{LLM-based models outperform non-LLM baselines.} Our explorations further advance this trend in AAC tasks, achieving SOTA results on both Clotho and AudioCaps.
    \item \textbf{Fine-tuned encoders matter.} Audio encoders like EAT, when fine-tuned on classification tasks, consistently outperform pre-trained or less optimized counterparts. 
    \item \textbf{Pre-training and PEFT enhance quality and efficiency.} Model pre-training combined with LoRA yields better captions with lower training cost.
    \item \textbf{RAG boost model's performance. } 
     By enabling access to external knowledge for describing unseen soundscapes, RAG enhances model's results.
    \item \textbf{Projection decoding narrows the modality gap. }To bridge the modality gap in CLAP, projection-based decoding proves effective, while direct use of the audio encoder in zero-shot settings performs unsatisfactorily. 
    \item \textbf{Even a single token can be sufficient.} In music captioning tasks, where information is relatively sparse, a strong encoder (MuQ) can effectively condense the relevant content into a single token and convey the content to the LLM decoder.
    \item \textbf{Optimal projector choice varies based on the task type.} Linear Projectors are preferred for tasks requiring strict monotonic alignment and temporal sequence preservation, such as ASR and VSR, while Q-Former bridges are more effective for tasks demanding global semantic perception and cross-modal context awareness, such as SEC and AAC. 
\end{itemize}
More detailed and technique-specific information can be found in sub-tasks research~\cite{chen2024slam, li2024drcap} of the SLAM-LLM series. 

\section{Conclusion}
SLAM-LLM presents a significant advancement in the field of multimodal large language models by providing a modular, flexible, and open-source framework specifically tailored for speech, language, audio, and music processing. 
Through its encoder–projector–LLM architecture, SLAM-LLM enables efficient customization and deployment across a wide range of tasks including ASR, SEC, AAC, and many other tasks. Experimental results demonstrate that recipes in SLAM-LLM not only achieve competitive or state-of-the-art performance on several benchmarks but also give insights for the LLM-based audio processing community. 
By lowering the entry barrier and promoting community collaboration, SLAM-LLM is poised to accelerate research and innovation in audio-language modeling and unlock new possibilities in multimodal AI. 

\bibliographystyle{IEEEtran}
\bibliography{custom}


 





\end{document}